# Deutsch and Jozsa's Algorithm Revisited

John W. Cooper*

IPST, University of Maryland, College Park MD, 20742

ABSTRACT

A classical analogue of Deutsch and Jozsa's algorithm is presented and its implications on quantum computing discussed.

Bob and Alice, longtime married CIA agents who worked in Boston and Washington respectively decided to get together for a weekend in New York so that they could share information without having their phone conversations overheard by Alice's mother Eve. Upon arrival in New York they checked into their hotel and proceeded to unpack. However, Alice noted that the dual light switches which turned on the light in the bathroom, one in the living room and one in the bathroom, were not working. She got a screwdriver out of her purse, removed the switch covers and found that although the wires in the living room were connected, those in the bath were not. "This is easy to fix" she said, and was about to connect the wires when Bob ( who has a doctorate in computer science from a famous British university) remarked, " Alice, we're professionals. Let's just call the desk and have one of the menials do the rewiring while we go out to lunch."

"OK", said Alice and placed a call to the desk. The desk assured her that the staff electrician, Mr. Oracle would be up immediately to fix the problem.

As they were finishing lunch Bob said "You know Alice, the problem of wiring those switches reminds me of Deutsch's first algorithm. If one must connect to both terminals on the left, there are only four ways in which it can be done." He then drew the following diagram on a paper napkin.

*E-mail address: jwc@umd.edu

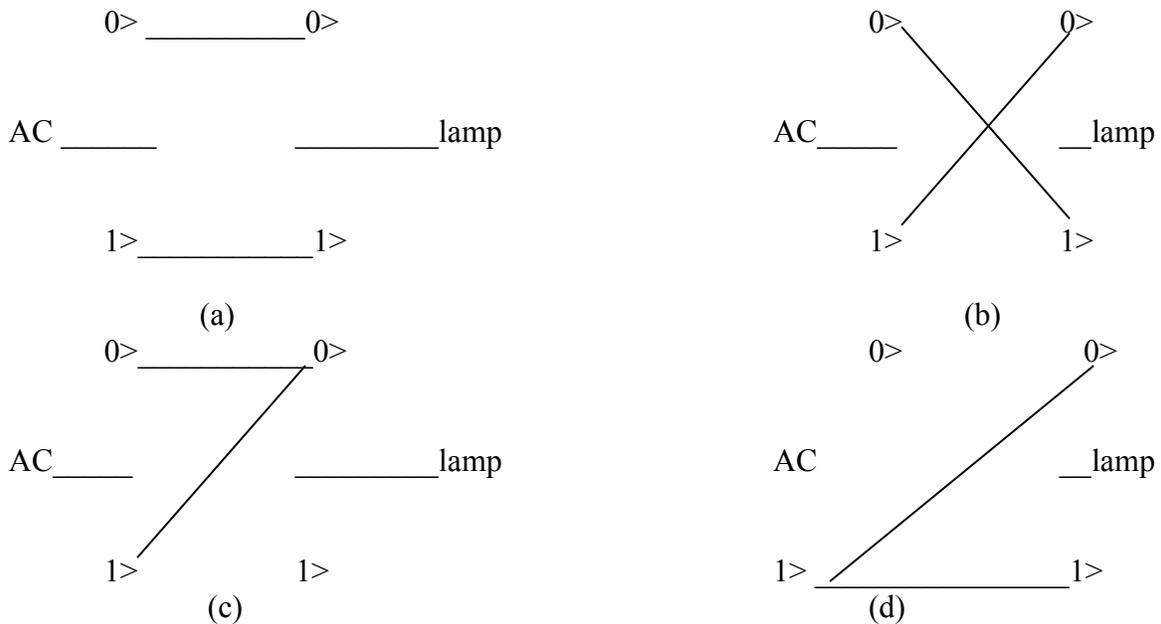

"If the wiring is done as in (a) the light will be on only when both switches are either in the up (0>) or down (1>) positions whereas for (b) the light will be on only if one switch is in the up (0>) and the other is in the down (1>) position. In (c) and (d) only the right switch will turn the light on and off and the left switch will have no effect. If we consider the switches to be bits (0> corresponding to "up" and 1> corresponding to "down") then (a) and (b) correspond to balanced mapping from left to right whereas (c) and (d) are constant mappings. The interesting thing is that in a classical system like this you must make two observations to determine whether the mapping is balanced or constant. Quantum mechanics allows you to determine this with just one observation."

"I don't understand" said Alice. "It seems to me that balanced is the only practical way to do the wiring so that both switches will work."

"Let me explain," said Bob and proceeded to draw the following truth table on his napkin.

| Left switch | Right switch | light on | light off |
| --- | --- | --- | --- |
| Up | Up | a or c | b or d |
| Down | Down | a or d | b or c |
| Up | Down | b or d | a or c |
| Down | Up | b or c | a or d |

"This table shows what you know about how the system is wired of each position of the switches. For example, if both switches are in the 'up' position and the light is on we know that the wiring corresponds to either a or c and if it is off the wiring is either b or d. None of the four possible settings of the switches will allow us to distinguish 'balanced' from 'constant'. To do this we must examine the system with two different switch settings and this corresponds to two separate observations. The marvelous thing about quantum mechanics is that it allows you to determine mappings like this with only one observation."

Alice thought about this for a moment and then burst out laughing. "That may be so" she said, "but I bet I can tell whether the wiring was done balanced or constant with one observation and I won't even bother to look at the light!"

"OK, it's a bet", said Bob and they paid for lunch and returned to the hotel.

As they entered their hotel room Mr. Oracle was just finishing the wiring and Alice remarked "I see you've finished the wiring and that you've done it correctly."

They then explained the bet they had made and Mr. Oracle agreed to redo the wiring several times and Alice would then indicate whether it was correct (i.e. balanced, either (a) or (b) or not. After about five or six re-wirings in which Alice correctly indicated whether the wiring was constant or balanced Bob became angry and made Alice wait in the hall while the re-wiring was done. It made no difference. Alice was correct every time. After 10 more tries they thanked Mr. Oracle for his efforts and he left.

Bob looked with wonder at Alice and said "OK, you win, but would you mind explaining how you knew each time?"

"It was easy", said Alice, "you see when I was in college, I worked part time as an electricians helper and we has some dumb electricians in the crew who would often wire that type of switch arrangement incorrectly (i.e., 'constant'). Since the left and right switches were often at some distance from each other and the power was not on yet, the inspectors designed a scheme that would insure that the wiring was done correctly. The rule was that when an electrician wired the second switch (i.e., the right one) he would always start with the switch in the up position and every time he connected a wire to the lower terminal he would flip the switch. That way all the inspector had to do was to see whether the switch was in the down position, meaning one wire was connected to the down terminal or in the up position, meaning that either two wires were connected to the down terminal or none were. I think they got the idea from you quantum computer people since they referred to the procedure as "f controlled not". At any rate they made every electrician promise to do this and it became standard procedure. I just assumed Mr. Oracle was following the same procedure and apparently he was."

The above allegory bears a relationship to what is known in the field of quantum computing as "Deutsch's XOR problem" (1) or "The first quantum algorithm (2). Physical realizations of this allegory have been made using a two qubit NMR computer (3,4). The algorithm has been cited as "the simplest possible example which illustrates the advantages of quantum computation through exploiting entangled states"(5). However it has been pointed out (6) that for the mapping of a single qubit onto another qubit there is no entanglement involved and the algorithm can be realized classically.

It is obvious that Bob and Alice view the wiring problem differently. Bob, with his background in computer science considers it as a simple case of function evaluation. The setting of the switch on the left corresponds to an input of either 0 or 1. After the wiring the output is either 0 or 1 which corresponds to the setting of the right hand switch. Bob is perfectly correct in stating that two settings of the switches are required to distinguish "balanced " from "constant" and this is obvious from the diagram. However, note that each constant case indicated in the truth table corresponds to effectively using only one of the wires in the diagram. Obviously, with only one wire being used one cannot define "balanced" and "constant".

Another interesting aspect of the allegory is that really, although there are four ways to do the wiring, from a practical standpoint there are really only two. As Alice points out, the circuit will work *only* if the wiring is balanced. Both wires are necessary.

Actually, the reason there are four different diagrams is that we have arbitrarily assumed that "up" corresponds to 0 and "down" to 1. With this convention rotating the right hand switch converts (a) to (b) and (c) to (d), ( or vice versa) without changing the circuit. Thus the unique separation of the four cases really depends upon adopting some convention as to which state corresponds to 0 and which to 1. The same will be true of a quantum mechanical system.

From the standpoint of the transfer of information the comparison of "balanced" and 'constant" is interesting. Note that for a constant mapping, the left hand switch has no effect at all ( although the right hand switch will work only if the left hand switch is in either the up or down position and not some where in between.) What this means is that in this simple case we obtain no quantitative information from a constant mapping.

For balanced mappings the two cases (a) and (c) correspond to either switching or not switching the labels of 0 and 1.

Perhaps the most interesting part of the allegory is Alice's explanation of how she knew how the wiring was done. The immediate reaction of a computer scientist to this would be to call foul since oracles are not supposed to devolve information as to how they performed transformations. However, as we shall see, Alice's evaluation of the situation corresponds rather closely to what is done it the "quantum mechanical" algorithm.

The mapping of the algorithm may be viewed in two ways; i.e., either as a mapping of one qubit to another, or as change in a given qubit as a result of the computation. In the first case the process is reversible since since the first qubit ( usually called x) is unaltered by the mapping and the initial state of the second qubit (f(x)) can be obtained simply by reversing the mapping process. On the other hand, if the mapping is interpreted as merely changing a single qubit, the process is irreversible for constant mapping. Alice's use of a second switch to determine whether the mapping is balanced or constant makes the mapping reversible in addition to solving the problem. The use of the second switch in Alice's explanation is similar to the use of a control bit in the usual interpretation of the algorithm. Two bits are necessary in order to make the process reversible.

The state of two qubits is often denoted as a two bit string consisting of two elements; i.e., x>y> where x> and y> are either 0 or 1. This is, of course, simply the tensor product and for two bits in the computational basis ( i.e. 0> and 1> for each qubit is equal to xy mod2 where an ordering of x and y is assumed. Alice's technique of switching the control switch when a given wire is connected to "1>" is equivalent to computing the tensor product of the initial and final states. The procedure works also if the designations 0> and 1> are reversed. Also, note that nothing is assumed about the amplitude of the initial switch settings. If we know nothing about the initial settings the amplitudes of each are $\pm\sqrt{2}/2$ and the probability of the switch being "up" or "down" is ½. Note that this assumes that the switch is in either "up" or "down". It cannot be in between.

The determination of a "constant" or "balanced" mapping has been compared to inserting a coin in a "black box" and determining by the output whether it is "fair" or "foul", i.e. heads on one side and tails on the other or either heads or tails on both sides(3). This does not seem to be a reasonable analogy. The algorithm corresponds more closely to actually changing a "fair" coin to "foul" (constant mapping), or either flipping the coin or leaving it it its initial orientation (balanced mapping).

It is also instructive to view the algorithm from the standpoint of parity. Parity is defined for a string of binary digits as even if the number of "1"s is devisable by 2 and odd otherwise. Quantum mechanically, the concept of parity for a single particle refers to whether or not its wave function changes sign when the sign of all of the coordinates upon which the wave function depends change sign. The concept is easily extended to many particle systems. A change of sign is denoted as odd parity and no change as even.

For a one bit binary system the parity is even if the bit is "0" and odd if it is "1". Thus the operation of changing the parity of a single bit simply refers to the action of a "not" gate.

For a qubit the situation is a little more complicated. Obviously, the classical definition does not apply since in general a qubit is in neither the 0> or 1> state. What is generally defined in quantum computing as parity is a property of a mathematical function of a single variable (7). Let x = 1,2,······N and consider a function f(x)= ± 1. Then the parity change is defined as as a difference in sign of a mapping of x to f(x).

With this definition, the flipping of the switch contingent on the mapping may be viewed as a change in parity. 0> → 1> and 1>→0> correspond to single changes of parity, 0>→ 0> and 1>→ 1> correspond to no change of parity and 0> → 1> and 1>→0> correspond to 2 changes of parity which means that the parity is unchanged. This is analogous to what happens when radiation interacts with a quantum mechanical system.  To the extent that the interaction is weak dipole radiation is the only important mechanism and this corresponds to a change in parity. In order for the parity to be unchanged , the radiation must either not interact with the system or must interact an even number of times since each interaction corresponds to a change of parity.

Introducing the concept of parity allows us to interpret our simple analogy in more general terms. Suppose, that instead of two wires , Mr. Oracle was faced with the problem of connecting N wires to the second switch. If the N wires consisted of two wire cables each of which was connected to a switch similar to the left switch he could connect each of the cables to the right side merrily flipping the switch as he worked in the same way he did for the first two wires. When he finished he would have connected N wires and N/2 cables and the position of the switch would indicate the parity of the mapping. The situation is completely analogous to that discussed in (7) and represents a classical solution to the parity problem without using a computer. It is also a solution of Deutsch's algorithm where the problem is to determine whether a function is "constant" or "balanced" for a large number of inputs.

After a leisurely dinner Bob and Alice resumed their discussion of the wiring problem and it's relationship to quantum computing. Alice remarked: "The concept of moving the switch was abandoned just after I left my job as electrician's helper. So nowadays they simply connect each wire to the lamp."  "But wouldn't that mean the lamp would be on all of the time regardless of the switch position?" asked Bob. "Yes, that is so," said Alice , "but after the wiring was done an electrical engineer arrived on the scene and inserted a device called a phase shifter in each of the wires, one on the left and one on the right. It seems someone realized that the power was alternating current and using the phase shifters instead of switches , one could dim the lights either inside the bathroom or in the living room using the phase shifters.  I thought it was a bad idea since I didn't want anyone dimming the light when I was reading in the bathroom. Besides, they found that there was a real problem in maintaining the correct setting on the phase shifters and they had to keep sending the engineer back constantly to correct the errors in the phases. I guess it provided work for people who otherwise would be unemployed, but I thought it was a bad idea."

The concept of using phase shifters which Alice mentioned is formally identical with the operation of a Mach – Zehnder optical interferometer, which is often used as an illustration of the power of quantum computing(2). In principle if it were possible to tune the wires a standing wave could be created along the 2 wire transmission line (8) and the relative phase in the two lines could be controlled. Alice is perfectly correct in that this is a bad idea and is pulling our leg in stating that this was done and made to work since this method of dimming lights is clearly impractical. However, dimming switches do exist and they rely on the fact that the power source is alternating current and do make use of the phase information of the power source. Phase correction is ,of course, one of the basic problems that has to be faced in any device that is intended to be used as an element of a quantum computer.

## References

1. R. Jozsa, "Quantum algorithms and the Fourier Transform." LANL preprint quant-phy 9707033.
2. R. Cleve, A. Ekert, L. Henderson, C. Macchiavello and M. Mosca, " ON Quantum Algorithms " LANL preprint quant phy 9903061
3. I. Chuang, L. Vandersypen, X. Zhou, D. Leung and S. Lloyd, "Experimental realization of a quantum algorithm", LANL preprint quant-phy 9801037.
4. A. Jones and M. Mosca, " Implementation of a Quantum Algorithm to Solve Deutsch's Problem on a Nuclear Magnetic Rosonance Quantum Computer", LANL preprint quant-phy 9801027.
5. V. Vedral and M. Plenio, Basics of Quantum Computation", LANL preprint Quant-phy 9802065.
6. D. Collins, K. Kim and W. Holton, "Deutsch-Jozsa algorithm as a test of quantum computation ", Phys Rev.A58,R1633 (1998).
7. E. Farhi, J Goldstone, S. Gutmann an M. Sipser, "Limit of the Speed of Quantum Computation in Determining Parity", Phys. Rev. Lttrs. 81 5542 (1998).
8. See for example "Communication Networks, Vol II "by E. A. Guillemin Wiley and Sons, New York 1935 p56.